\documentclass[%
 reprint,
 amsmath,amssymb,
 aps,
prb,
]{revtex4-1}
\usepackage[utf8]{inputenc}
\usepackage{graphicx}
\usepackage{dcolumn}
\usepackage{bm}
\usepackage{hyperref}
\usepackage{amsmath}
\usepackage{amssymb}
\usepackage{pdfpages}
\makeatletter
\AtBeginDocument{\let\LS@rot\@undefined}
\makeatother

\begin{document}
\title{Gate-Defined One-Dimensional Channel and Broken Symmetry States in MoS$_2$ van der Waals Heterostructures }

\author{Riccardo Pisoni}

\author{Yongjin Lee}
\author{Hiske Overweg}
\author{Marius Eich}
\author{Pauline Simonet}
\affiliation{%
Solid State Physics Laboratory, ETH Zürich,~CH-8093~Zürich, Switzerland}

\author{Kenji Watanabe}
\author{Takashi Taniguchi}
\affiliation{National Institute for Material Science, 1-1 Namiki, Tsukuba 305-0044, Japan}
\author{Roman Gorbachev}
\affiliation{%
National Graphene Institute, University of Manchester,
Manchester M13 9PL, United Kingdom}
\author{Thomas Ihn}
\author{Klaus Ensslin}
\affiliation{%
Solid State Physics Laboratory, ETH Zürich,~CH-8093~Zürich, Switzerland}

 \email{pisonir@phys.ethz.ch}

\begin{abstract}
 We have realized encapsulated trilayer MoS$_2$ devices with gated graphene contacts. In the bulk, we observe an electron mobility as high as 7000~cm$^{2}$/(V s) at a density of 3 $\times$ 10$^{12}$~cm$^{-2}$ at a temperature of 1.9~K. Shubnikov--de Haas oscillations start at magnetic fields as low as 0.9~T. The observed 3-fold Landau level degeneracy can be understood based on the valley Zeeman effect. Negatively biased split gate electrodes allow us to form a channel that can be completely pinched off for sufficiently large gate voltages. The measured conductance displays plateau-like features.
\end{abstract}
\maketitle


Laterally confined two-dimensional (2D) materials offer the opportunity to engineer quantum states with tunable spin, charge and even valley degrees of freedom\cite{hanson_spins_2007,loss_quantum_1998,petta_coherent_2005}. The pure thinness of these materials in combination with 2D insulators such as boron nitride pave the way for ultrasmall strongly coupled gate-defined quantum devices\cite{goossens_gate-defined_2012,song_gate_2015,wang_engineering_2016,novoselov_two-dimensional_2005}. In addition the variety of transition metal dichalcogenides (TMDCs) materials will allow to choose different strength of spin-orbit interaction that is relevant for electric control of spin and valley-states in view of quantum information processing. In this Letter, we describe a split gate geometry realized on a high-quality molybdenum disulfide (MoS$_2$) van der Waals heterostructure that results in a tunable tunneling barrier, the starting point for any electronic quantum device. The electronic quality of our trilayer MoS$_2$ device is documented by the observation of Shubnikov-de Haas oscillations (SdHO) occurring at magnetic fields as low as 0.9 T. In addition a 3-fold degeneracy of the Landau levels (LLs) is observed arising from the 3 Q and 3 Q' valleys situated in the middle of the Brillouin zone and shifted in magnetic field by the valley Zeeman effect\cite{li_valley_2014,srivastava_valley_2015,macneill_breaking_2015,aivazian_magnetic_2015,wu_evenodd_2016}. The constriction can be completely pinched off with resistances values exceeding the quantum of resistance $h/e^{2}$ by orders of magnitude, a prerequisite for the realization of any single-electron transistor. We observe the occurrence of plateau-like features in the conductance with a spacing of the order of $e^{2}/h$. These experiments are a first step toward gate controlled quantum devices in transition metal dichalcogenides.

\begin{figure*}
\centering
\includegraphics[width=\textwidth]{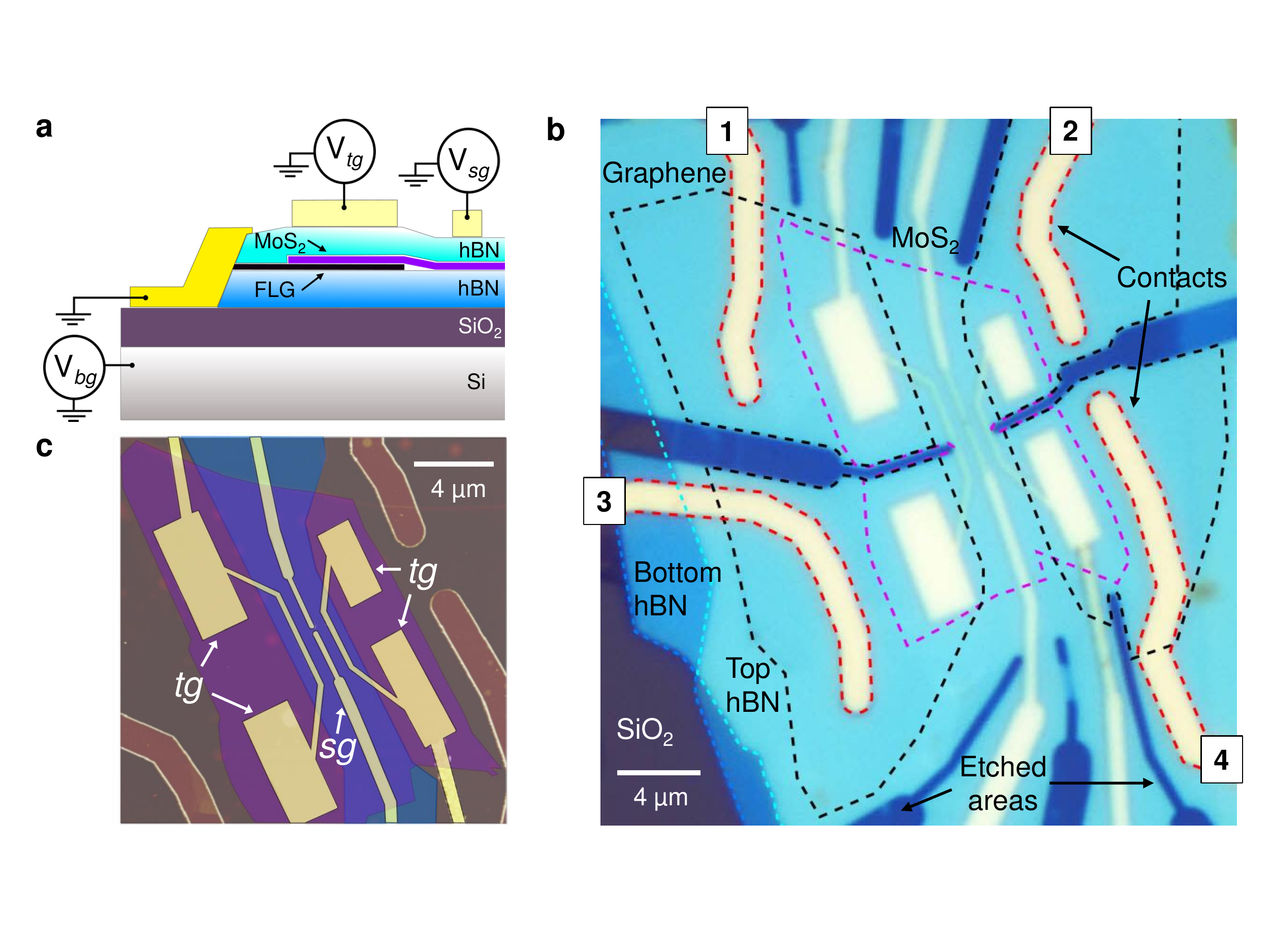}
\caption{(a) Cross-sectional schematic of the MoS$_2$-based field-effect device. (b) Optical micrograph of the device. An encapsulated trilayer MoS$_2$ (purple dashed lines) is connected to two graphene flakes (black dashed lines). Au/Cr one-dimensional edge contacts (numbered 1-4) to the graphene flakes are fabricated\cite{wang_one-dimensional_2013}. The dark blue regions outline the etched areas that define the final device geometry. (c) False-color AFM image of the device before the last etching process. Four rectangular gates have been deposited on top of the contact areas between graphene and MoS$_2$ in order to reduce the contact resistance without affecting the low carrier density in the MoS$_2$ channel. A split gate, defining a nanoconstriction, has been placed on top of a bubble-free region.
}
\label{fig:1}
\end{figure*}

To achieve high mobility TMDC devices, we fabricate MoS$_2$-based van der Waals heterostructures. As shown schematically in Figure~\ref{fig:1}a, a trilayer MoS$_2$ flake ($\sim$ 2~nm thick), contacted with two few-layer graphene (FLG) sheets, is encapsulated between hexagonal boron nitride (hBN) crystals\cite{lee_highly_2015,cui_multi-terminal_2015,liu_toward_2015}. The bottom one is 30 nm thick and separates the MoS$_2$ from substrate phonons and charged impurities, further serving as an atomically flat substrate\cite{xue_scanning_2011}. The top one is 20 nm thick and prevents the adsorption of organic residues during the fabrication process. To assemble the heterostructure we employ a polymer-based dry pick-up and transfer technique\cite{wang_one-dimensional_2013,wang_electronic_2015} using a polycarbonate film\cite{zomer_fast_2014,bretheau_tunnelling_2017} supported by polydimethylsiloxane. Assembling and exfoliating the various thin films was performed in an argon environment\cite{cao_quality_2015}. The films' thicknesses were first determined from the optical contrast and then verified by atomic force microscopy (AFM). The top hBN crystal serves as the dielectric layer for the top gates whereas the SiO$_2$/n-Si substrate works as the bottom dielectric (285~nm) and gate electrode. The top gate structure consists of two pairs of local gates on top of the contact areas between graphene and MoS$_2$, and a split gate with a 100~nm gap (Figure~\ref{fig:1}c). In order to avoid electrostatic inhomogeneities, the split gate has been deposited on a bubble-free area on top of the MoS$_2$ channel, found by AFM. Cr/Au electrodes are used as one-dimensional edge contacts to graphene (red dashed lines in Figure~\ref{fig:1}b)\cite{wang_one-dimensional_2013}. Finally, the four-terminal device geometry is defined by etching through the top hBN, MoS$_2$ and FLG (Figure~\ref{fig:1}b). As shown in Figure~\ref{fig:1}a, we can tune the carrier density in the MoS$_2$ channel by biasing the Si gate electrode ($V_\mathrm{bg}$). We can further tune locally the carrier density in the contact area between MoS$_2$ and graphene ($V_\mathrm{tg}$) and deplete the MoS$_2$ two-dimensional electron gas (2DEG) ($V_\mathrm{sg}$).

\begin{figure*}
\centering
\includegraphics[width=0.5\textwidth]{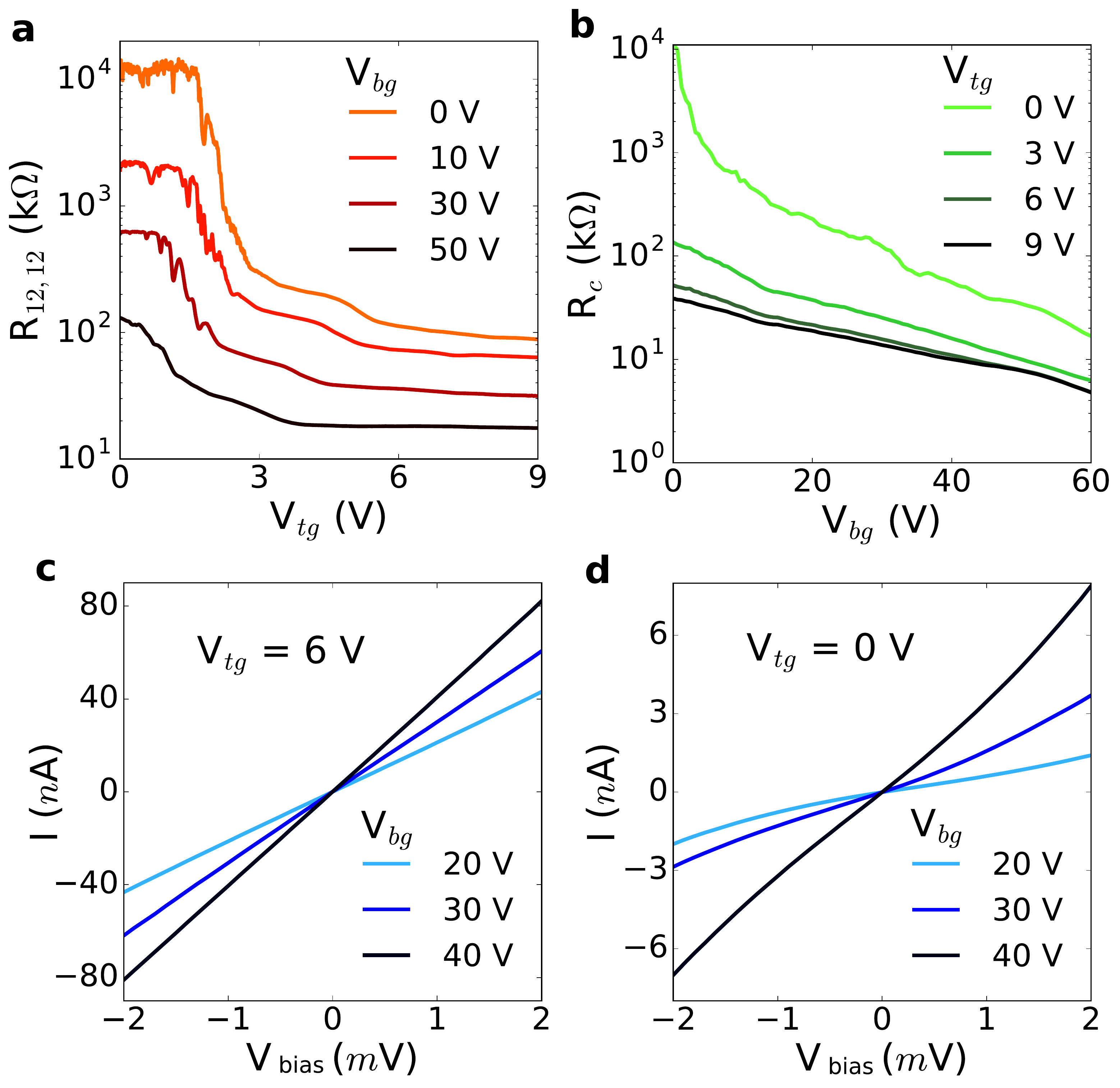}
\caption{(a) Two-terminal resistance as a function of $V_\mathrm{tg}$ at different $V_\mathrm{bg}$, $R_{12,12}$, refers to contact numbering in Figure~\ref{fig:1}b, $T = 4$~K. The resistance decreases by up to 2 orders of magnitude by biasing the gates on top of the contact area between graphene and MoS$_2$. The effect of the top gates decreases with increasing $V_\mathrm{bg}$  due to the flattening of the MoS$_2$ conduction band at the interface. The resistances stop decreasing above $V_\mathrm{tg} = 5$~V, where the MoS$_2$ sheet resistance and the residual contact resistance between MoS$_2$ and graphene persist. (b) Contact resistance as a function of $V_\mathrm{bg}$ at different $V_\mathrm{tg}$. Biasing the top gates decreases the contact resistance by up to 3 orders of magnitude. (c) Linear $I -V_\mathrm{bias}$ behavior is observed when $V_\mathrm{tg} > 3$~V. (d) With $V_\mathrm{tg} = 0$~V, nonlinear $I -V_\mathrm{bias}$  behavior is observed. 
}
\label{fig:2}
\end{figure*}

Figure~\ref{fig:2}a shows the low temperature ($T = 4$~K) two-terminal resistance as a function of $V_\mathrm{tg}$, for different $V_\mathrm{bg}$. The resistance decreases with increasing $V_\mathrm{bg}$, as expected for an n-type semiconductor\cite{cui_multi-terminal_2015}. At fixed $V_\mathrm{bg}$, the resistance drops for increasing $V_\mathrm{tg}$ by up to 2 orders of magnitude. Figure~\ref{fig:2}b displays the estimated contact resistance $R_{c} = \frac{1}{2}(R_{12,12}-R_{12,34})$, where $R_{12,12}$ is the two-probe resistance and $R_{12,34}$ is the four-probe resistance of MoS$_2$ (Figure~\ref{fig:1}b). $R_{c}$ can be improved by up to 3 orders of magnitude with increasing $V_\mathrm{tg}$. Figure~\ref{fig:2}c,d shows the current flowing into the device ($I$) as a function of the voltage applied between two graphene electrodes ($V_\mathrm{bias}$). When $V_\mathrm{tg} = 0$~V, non linear $I -V_\mathrm{bias}$ curves are observed indicating gapped-behavior corresponding to nonohmic contacts (Figure~\ref{fig:2}d). Linear behavior is achieved at higher $V_\mathrm{bg}$ as already demonstrated in previous works\cite{lee_highly_2015,cui_multi-terminal_2015,liu_toward_2015}. We observe linear $I -V_\mathrm{bias}$ curves at any  $V_\mathrm{bg} \geq 0$~V when $V_\mathrm{tg} > 3$~V (Figure~\ref{fig:2}b).  Therefore, we can locally tune the carrier density in the MoS$_2$ layer and the Fermi level of graphene to achieve ohmic contact without compromising the low carrier density in the MoS$_2$ channel. This allows us to investigate the physics of MoS$_2$ at the edge of the conduction band.

\begin{figure*}
\centering
\includegraphics[width=\textwidth]{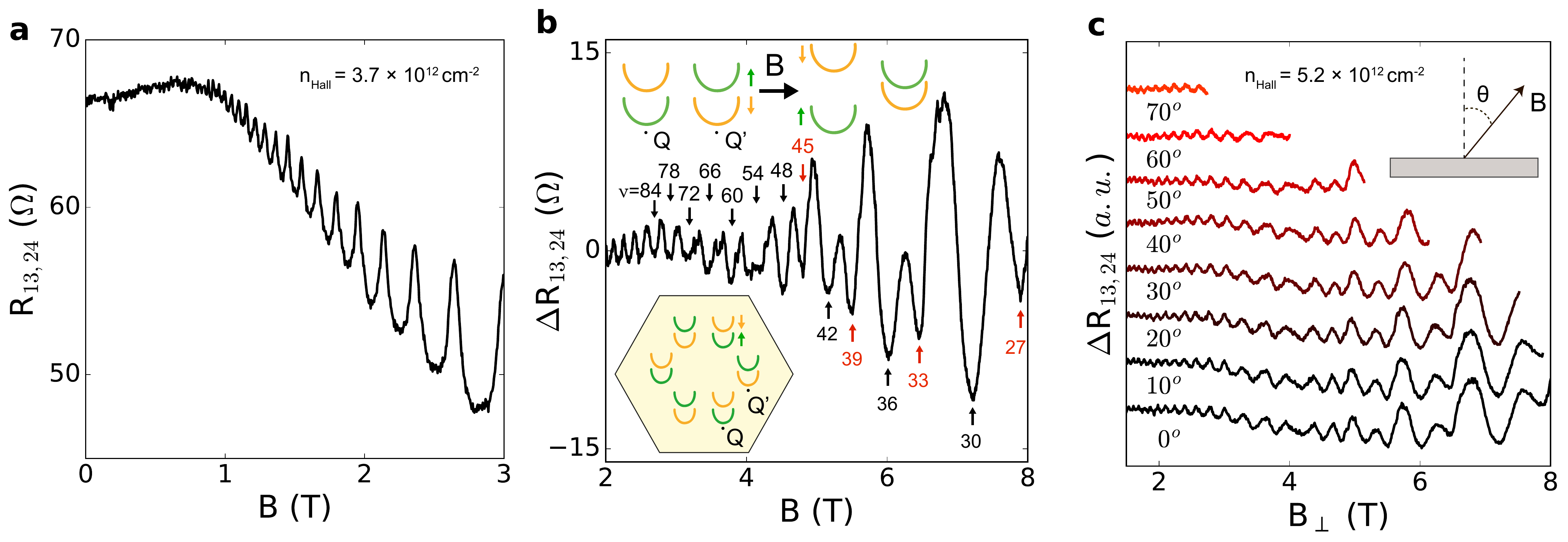}
\caption{(a) Four-terminal resistance as a function of B field for a 3L MoS$_2$ device measured at $T = 1.9$~K; $n = 3.7\times10^{12}$~cm$^{-2}$. (b) Background subtracted four-terminal resistance as a function of $B$ field at $n = 5.2\times10^{12}$~cm$^{-2}$. The LL filling factors are labeled for the oscillation minima. The degeneracy of 6 arises from the degeneracy of the 3 Q and 3 Q’ valleys in the conduction band (black arrows). Because of time reversal symmetry at $B = 0$~T  and broken inversion symmetry, the spin degeneracy within each Q or Q’ valley is already lifted. The degeneracy of 6 can be lifted at relatively high magnetic field due to the valley Zeeman effect (red arrows). Insets: Schematic diagrams for the Bloch bands of 3L MoS$_2$ . When $B=0$~T, Q and Q’ valleys are degenerate, spin up is displayed in green, spin down in orange. When $B>0$~T, the valley Zeeman effect lifts the degeneracy. (c) SdHO shown in (b) as a function of $B_{\perp}$ at different tilted angles. The traces are shifted vertically for clarity. Inset: schematic of the sample orientation with respect to the magnetic field $B$.
}
\label{fig:3}
\end{figure*}

To examine the quality of our device we performed magnetotransport measurements at $T = 1.9$~K. Four-probe measurements, using the standard lock-in technique at 80.31 Hz, can be performed due to the reasonably good ohmic contacts. Figure~\ref{fig:3}a shows the four-terminal resistance $R_{13,24}$ as a function of the magnetic field $B$, at $n=3.7\times10^{12}$~cm$^{-2}$ . We observe SdHO starting at $B \approx 0.9$~T, which yields a lower boundary for the quantum mobility of about 11000 cm$^{2}$/Vs, in agreement with the measured van der Pauw mobility of 7000 cm$^{2}$/Vs. Figure~\ref{fig:3}b displays four-terminal resistance $\Delta$$R_{13,24}$ with a smooth background subtracted, as a function of $B$, at $n=5.2\times10^{12}$~cm$^{-2}$. Distinctive features appear above $B\approx 4$~T on top of the SdHO. They first emerge as shoulder-like features developing then into local minima in the SdHO as shown in Figure~\ref{fig:3}b with red arrows. From the SdHO we can determine the density of the 2DEG, $n=(de/h)(1/\Delta(1/B))$ where $\Delta(1/B)$ is the period of the SdHO and the prefactor $d$ accounts for spin and valley degeneracies. The electron density calculated from SdHO matches the Hall density when $d = 6$. As shown in Figure \ref{fig:3}b, a 6-fold LL degeneracy is clearly observed at relatively low magnetic field. Above $4$~T additional minima for filling factors $\nu=$~27, 33, 39 and 45 appear. The degeneracy of 6 arises from the 3 Q and 3 Q' valleys located along $\Gamma-K$ symmetry lines in the first Brillouin zone (inset of Fig.~\ref{fig:3}b), which correspond to the 6 degenerate conduction band minima expected from the band structure calculations of trilayer MoS$_2$\cite{cheiwchanchamnangij_quasiparticle_2012,wang_electronics_2012,mak_atomically_2010,cappelluti_tight-binding_2013,chhowalla_chemistry_2013,wu_evenodd_2016}. The spin degeneracy within each Q and Q' valley is already lifted by broken inversion and time reversal symmetry. At relatively high magnetic fields, due to the opposite spin character at Q and Q' valley in 2D TMDC\cite{xiao_coupled_2012,chhowalla_chemistry_2013}, the Zeeman splitting becomes comparable to the LL splitting and we observe the LL sextet being lifted into two LL triplets (inset of Figure~\ref{fig:3}b). Figure~\ref{fig:3}c shows $\Delta$$R_{13,24}$ as a function of the magnetic field component perpendicular to the 2DEG plane ($B_{\perp}$) at different angles $\theta$ (inset of Figure~\ref{fig:3}c). The SdHOs remain unchanged for all values of $\theta$ up to $\theta=$~70$^{\circ}$. A similar behavior was observed for mono- and bilayer WSe$_2$\cite{movva_density-dependent_2017}. The insensitivity of the Zeeman energy to the parallel component of the magnetic field may indicate that the electron spin is locked perpendicular to the plane due to the strong spin-orbit coupling and broken inversion symmetry in trilayer MoS$_2$\cite{wu_evenodd_2016}.

\begin{figure*}
\centering
\includegraphics[width=\textwidth]{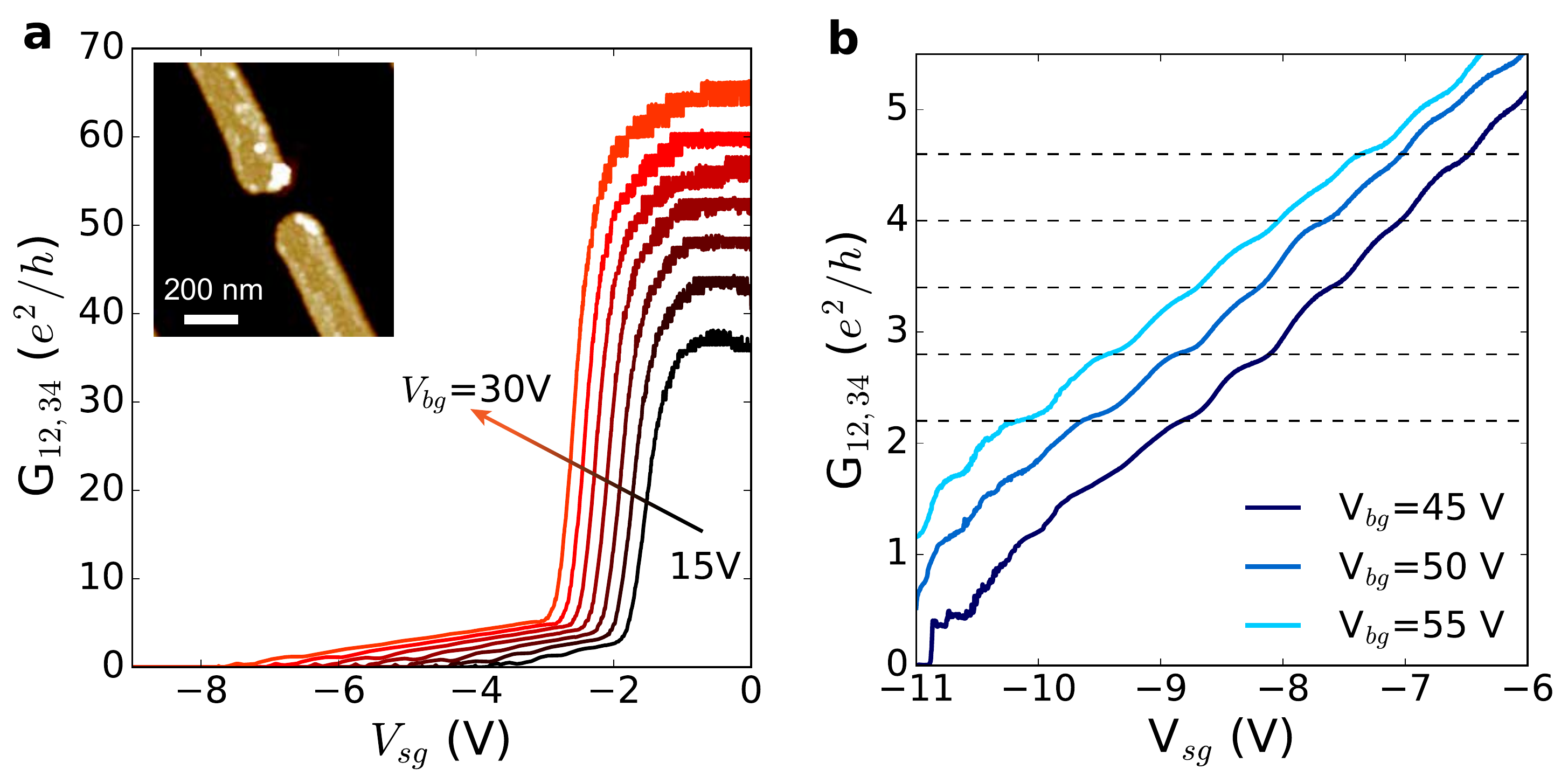}
\caption{(a) Evolution of  four-terminal conductance pinch off curves as a function of $V_\mathrm{bg}$ at $T = 1.9$~K. The QPC channel can be pinched off over a large range of electron density.
Inset: AFM micrograph of the QPC split gate. The opening is 100~nm wide and 200~nm long. (b) Four-terminal conductance as a function of $V_\mathrm{sg}$ at various $V_\mathrm{bg}$. Plateau-like features, marked by horizontal dashed lines, appear with a spacing compatible with $e^{2}/h$.
}
\label{fig:4}
\end{figure*}

High-quality MoS$_2$ 2DEGs allow us to investigate quantum transport in gate-defined nanostructures. As shown in the inset of Figure~\ref{fig:4}a, we define a constriction, 100~nm wide and 200~nm long, by electrostatically depleting the MoS$_2$ layer. In  Figure~\ref{fig:4}a, we display the measured four-terminal conductance $G_{12,34}$ in unit of the conductance quantum,\cite{wharam_one-dimensional_1988,van_wees_quantized_1988} $e^{2}/h$, as a function of $V_\mathrm{sg}$ and $V_\mathrm{bg}$, at $T = 1.9$~K. At higher electron density, that is, higher $V_\mathrm{bg}$, a more negative $V_\mathrm{sg}$ is required in order to pinch off the  MoS$_2$ channel with on-off ratios exceeding $10^{5}$. Figure~\ref{fig:4}b shows $G_{12,34}$, close to pinch-off, at different $V_\mathrm{bg}$. At relatively high carrier density ($n \approx 5\times10^{12}$~cm$^{-2}$, Figure~\ref{fig:4}b), we observe a significant decrease of resonances close to pinch-off, which we attribute to localized states forming in the electrostatically confined channel. These states, caused by the disorder potential in the MoS$_2$, may be better screened at high carrier density. At $B = 0$~T, we observe even-spaced plateau-like features within 2 and 5 $e^{2}/h$. While plateau-like features are expected at multiples of 6e$^{2}$/h, the experimental results show features that are roughly spaced by e$^{2}$/h, and even these values are not met precisely. Level degeneracies might be lifted by the additional confinement of the quantum point contact. However, we do not know the exact reason why the experimental data show plateau-like features that match only qualitatively a quantization with values around multiples of e$^{2}$/h. Further improvement of the 2DEG quality is required to demonstrate exact conductance quantization. Figure~\ref{fig:4}b shows raw data for the conductance. Subtracting a series resistance to account for possible contributions of the surrounding areas of the electron gas did not lead to a better matching of the plateau-like features with the expected values of the conductance quantization.

In conclusion, we have developed a van der Waals heterostructure platform that allows us to obtain high-quality 2DEGs in MoS$_2$, displaying an electron mobility of $7000$ cm$^{2}$/(V s) with electron density as low as $\sim$ 10$^{12}$~cm$^{-2}$. We observe SdHO starting at magnetic fields as low as 0.9 T with a 6-fold LL degeneracy that is lifted into a 3-fold LL with magnetic field. We further observe signatures of quantized conductance by electrostatically depleting a split gate on the MoS$_2$ 2DEG. The realization of an electrostatically tunable tunneling barrier reaching full pinch-off is the first step toward gate-defined quantum dots in 2D semiconducting TMDC in order to control and manipulate the spin and valley states of single confined electrons.\cite{novoselov_2d_2016,kormanyos_spin-orbit_2014,loss_quantum_1998}

During preparation of the manuscript, we became aware of related works.\cite{epping_quantum_2016,wang_engineering_2016}

\section*{acknowledgement}
We thank Guido Burkard,  Andras Kis, Matija Karalic and Christopher Mittag for fruitful discussions. We acknowledge financial support by the Graphene Flagship, the EU Spin-Nano RTN network, and by the National Center of Competence in Research on Quantum Science and Technology (NCCR QSIT) funded by the Swiss National Science Foundation. Growth of hexagonal boron nitride crystals was supported by the Elemental Strategy Initiative conducted by the MEXT, Japan and JSPS KAKENHI Grant Numbers JP15K21722.

\bibliography{arXiv}

\end{document}